\documentclass[sn-mathphys-num]{sn-jnl}%
\usepackage{graphicx}%
\usepackage{multirow}%
\usepackage{amsmath,amssymb,amsfonts}%
\usepackage{amsthm}%
\usepackage{mathrsfs}%
\usepackage[title]{appendix}%
\usepackage{xcolor}%
\usepackage{textcomp}%
\usepackage{manyfoot}%
\usepackage{booktabs}%
\usepackage{algorithm}%
\usepackage{algorithmicx}%
\usepackage{algpseudocode}%
\usepackage{listings}%

\theoremstyle{thmstyleone}%
\theoremstyle{thmstyletwo}%

\theoremstyle{thmstylethree}%
\begin{document}

\title[Article Title]{Direct Measurement of Dark Matter Mass through the Scattering of a High-Energy Matter Beam}

\author*[1]{\fnm{Bibhabasu} \sur{De}}\email{bibhabasude@gmail.com}



\affil*[1]{\orgdiv{Department of Physics}, \orgname{The ICFAI University Tripura}, \orgaddress{\city{Kamalghat}, \postcode{799210}, \state{Tripura}, \country{India}}}


\abstract{The traditional methods of estimating the Dark Matter~(DM) mass scale crucially depend on the assumptions about the interaction mechanism between the DM and the Standard Model~(SM) sectors, making it challenging to achieve precise mass measurements. However, in case of a successful scattering event of a high-energy matter beam on a DM particle, the DM mass can be directly measured without probing the {\it New Physics}~(NP) interaction. The present chapter discusses a model-independent kinematical analysis to formulate the DM mass as a function of experimentally observable quantities. The approach can be significant to precisely pin-down the exact mass value of the interacting DM particle and offers insights into the DM-SM interaction strength.}




\maketitle

\section{Introduction}\label{sec1}

The existence of DM has already been established through various cosmological~\cite{Planck:2018vyg} and astrophysical observations~(for a recent review, see Ref.~\cite{Cirelli:2024ssz}). However, to reveal its fundamental characteristics~(i.e., mass, spin, charge), it's necessary to probe the non-gravitational interaction(s) of the DM. In literature, there exist several theoretical formulations predicting an experimentally viable particle DM, e.g., weakly interacting massive particle~(WIMP)~\cite{Jungman:1995df}, feebly interacting massive particle~(FIMP)~\cite{Hall:2009bx}, weakly interacting slim particle~(WISP)~\cite{Arias:2012az}, strongly interacting massive particle~(SIMP)~\cite{Hochberg:2014dra}, etc. However, till now, only null results have been reported from the experiments. 

The conventional direct detection~(DD) experiments~(for a review, see Ref.~\cite{Misiaszek:2023sxe}) rely upon the scattering of a DM particle on the regular matter. The Standard Model~(SM) states~(nucleus or electron) act as the target, and the DM is expected to interact non-gravitationally as the earth passes through the galactic halo. For the WIMPs, nuclear recoil experiments set the most stringent bounds~\cite{LZ:2022lsv, XENON:2020kmp, PandaX:2024qfu}, whereas in the sub-GeV mass regime DM-electron scattering~\cite{Essig:2011nj} may result in a more promising signal~\cite{XENON:2019zpr, DarkSide:2022knj,PandaX-II:2021nsg,SENSEI:2023zdf}. The fundamental principle behind all the existing DD experiments is to search for the deposited energy, which may potentially come from the kinetic energy of the incoming DM. If that exceeds the detector threshold, the event will be observed. However, the energy deposition depends on two factors: $(i)$ the effective interaction strength between the DM~($\chi$) and the SM~($B$) fields, and $(ii)$ the kinematics of the DM-target 2-body system. Thus, for a fixed detector sensitivity, the interaction cross section~($\sigma_{\chi B}$) can only be constrained as a function of the DM mass $m_\chi$. Therefore, following the philosophy of the present DD experiments, it is impossible to determine the exact mass of the DM particle without prior information on the underlying NP that connects the DM and the SM sectors. However, very recently, Refs.~\cite{Ruzi:2023mxp, Yu:2024spj} have proposed a DD experiment from a slightly different perspective where the DM particle acts as the target and high-energy muons are used to probe the NP interaction. For a $\mu$-philic DM, a scattering event might result in a non-zero scattering angle~($\theta$) between the incoming and the outgoing muon beams. In the proposed detector, the number of events has been simulated as a function of $\cos\theta$ for different DM masses. Thus, the recoil energy can be formulated as a function of the scattering angle leading to a projected bound on $\sigma_{\chi\mu}$ for $m_\chi\geq \mathcal{O}(m_\mu)$.

This chapter adopts the idea in a generic form and proposes a simple model-independent method for the experimental determination of DM mass within the framework of the DD searches. The current DD experiments are mainly designed to measure the recoil energy of the SM particle. For the nuclear targets, the entire energy deposited by the DM appears as the recoil energy of the nucleus, whereas for the atoms, a fraction of the net available energy is used to overcome the binding energy of the electrons, and the rest triggers the recoil. Due to the directional uncertainty of the incoming DM particle, the angular distribution of the recoiled state becomes insignificant in the DD experiments. However, following the proposal of the PKU-muon experiment~\cite{Yu:2024spj}, with a quasi-static DM target, it is possible to trace the angular position of the scattered matter beam, resulting in a direct measurement of $\cos\theta$ with a high degree of precision. Through a simple kinematical analysis, it has been shown here that significant information on the DM mass can be extracted if the proposed detector were improvised to record the energy of the outgoing beam as well. Further, the results can used to explore the NP that governs the $\chi-B$ interaction.

The rest of the chapter is organized as follows. Using the basic kinematics of a 2-body scattering, Sec.~\ref{sec2} formulates the DM mass as a function of $\cos\theta$ and the final state energy of the beam particle with Sec.~\ref{sec3} presenting a brief summarization of the work. Note that, for the subsequent discussions, natural units will be used, i.e., $\hbar=1$ and $c=1$ will be considered for all the analytical calculations.

\section{Scattering on a Dark Target}\label{sec2}

\begin{figure}[h]
 \centering
 \includegraphics[scale=0.45]{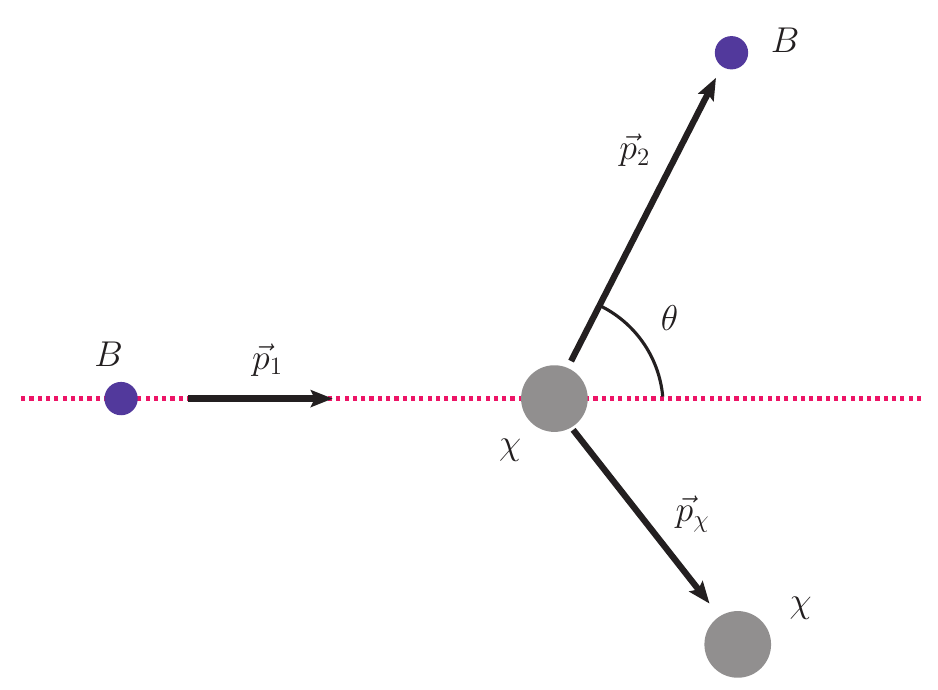}
 \caption{Scattering of the SM field $B$ on the dark matter particle $\chi$.}\label{fig:scat}
 \end{figure}
 
Fig.~\ref{fig:scat} shows the scattering of a SM particle on the DM state $\chi$. Note that $B$ generically symbolizes any possible SM field that may have a weak or feeble interaction with $\chi$. Note that, though the DM is not at rest with respect to the earth, for the high-energy matter beams, $\chi$ can be considered a quasi-static target. Thus, the relative velocity $|\vec{v}_B-\vec{v}_\chi|\approx |\vec{v}_B|$ and one can assume the rest frame of $\chi$ being approximately coincident with the earth frame of reference.
 
  Therefore, the momentum conservation leads to 
 \begin{align}
 &~\vec{p}_1=\vec{p}_2+\vec{p}_\chi\nonumber\\
 \Rightarrow &~ |\vec{p}_\chi|^2=|\vec{p}_1|^2+|\vec{p}_2|^2-2|\vec{p}_1||\vec{p}_2|\cos\theta,
 \label{eq:px}
 \end{align}
 where $\vec{p}_1$ and $\vec{p}_2$ denote the momentum of the incoming and outgoing matter beams, respectively. $ \vec{p}_\chi$ is the momentum of the scattered DM state, with $\theta$ being the scattering angle of the matter particle. For an elastic scattering between the matter particle and $\chi$, energy conservation results in
 \begin{align}
 &~E_1+m_\chi=E_2+\left(m^2_\chi+|\vec{p}_\chi|^2\right)^{1/2}\nonumber\\
 \Rightarrow &~|\vec{p}_\chi|^2=(E_1-E_2)\Big[E_1-E_2+2m_\chi\Big].
 \label{eq:px1}
 \end{align}
 Comparing Eqs.~\eqref{eq:px} and \eqref{eq:px1}, one obtains,
 \begin{align}
 &~(E_1-E_2)\Big[E_1-E_2+2m_\chi\Big]=|\vec{p}_1|^2+|\vec{p}_2|^2-2|\vec{p}_1||\vec{p}_2|\cos\theta\nonumber\\
 \Rightarrow&~(E_1-E_2)^2+2m_\chi(E_1-E_2)=E_1^2-M_B^2+E_2^2-M_B^2-2|\vec{p}_1||\vec{p}_2|\cos\theta\nonumber\\
 \Rightarrow&~-2E_1E_2+2m_\chi(E_1-E_2)=-2M_B^2-2|\vec{p}_1||\vec{p}_2|\cos\theta\nonumber\\
 \Rightarrow&~E_1E_2-m_\chi(E_1-E_2)=M_B^2+\Big\{\left(E_1^2-M_{B}^2\right)\left(E_2^2-M_{B}^2\right)\Big\}^{1/2}\cos\theta\nonumber\\
  \Rightarrow&~m_\chi=\left(\frac{1}{E_1-E_2}\right)\Bigg[E_1E_2- M_{B}^2-\Big\{\left(E_1^2-M_{B}^2\right)\left(E_2^2-M_{B}^2\right)\Big\}^{1/2}\cos\theta\Bigg].
 \label{eq:mx1}
 \end{align}
 Here, $m_\chi$ and $M_B$ are the rest masses of $\chi$ and the beam particle, respectively. In the high-energy limit, i.e., $E_1,\,E_2\gg M_{B}$, Eq.~\eqref{eq:mx1} reduces to,
 \begin{align}
 m_\chi=&~\left(\frac{E_1E_2}{E_1-E_2}\right)(1-\cos\theta)\nonumber\\
 =&~\frac{E_2\,(1-\cos\theta)}{1-E_2/E_1}~.
 \label{eq:mass}
\end{align}  
Note that, in principle, $E_2$ and $\cos\theta$ are correlated through the NP interaction; $\cos\theta\to 1$ physically implies $E_2/E_1\to 1$, i.e., there is no scattering. This is the situation of either being in a DM-void region of the galaxy or having a negligibly small~(beyond the detector sensitivity) effective interaction between $\chi$ and the considered SM field. However, in case of a notable scattering event, if a detector can record the angular distribution and the energy deposition of the scattered beam as two independent parameters, Eq.~\eqref{eq:mass} can be significant to determine the exact mass of the scatterer~(i.e., the DM particle) without requiring explicit knowledge of the interaction cross section. Further, it is worth mentioning that with precise information of $E_2$~(or equivalently $E_1-E_2$) and $m_\chi$, one can accurately determine the scattering cross section and, hence, estimate the effective NP interaction strength.
\section{Conclusion}\label{sec3}
The discussion follows from a very recent proposal~(PKU-muon experiment) of using high-energy muon beams to probe the DM-SM interaction. With respect to an ultrarelativistic matter beam, the DM can be considered a quasi-static target, and one can use the fundamental kinematics of 2-body elastic scattering to formulate the DM mass as a function of the energy and scattering angle of the outgoing SM state. The proposed PKU setup is currently planning to measure only the angular distribution of the scattered muon beam/cosmic muons leading to a projected bound on the $\sigma_{\chi\mu}$ as a function of $m_\chi$. The present work proposes a minimal improvisation of the detector principle and shows that with a non-trivial energy value and angular deflection of the scattered matter beam, $m_\chi$ can be exactly measured. The final energy and scattering angle of the SM particle must be noted as two independent parameters to eliminate the requirement of probing NP interactions for direct determination of the DM mass. However, the proposal can only succeed if $\chi$ is significantly abundant within the detector and possesses a non-negligible interaction with the beam particle.


\bibliography{DM_Mass_SN}

\end{document}